\documentclass[intlimits,twoside,a4paper]{article}

\usepackage{amsmath,amssymb}
\usepackage{graphicx}
\usepackage{wrapfig}

\usepackage[T2A]{fontenc}
\usepackage[cp1251]{inputenc}

\usepackage[eqsecnum]{cmpj2}

\issue{2015}{18}{4}{43603}
\doinumber{10.5488/CMP.18.43603}

\title[Stokes-Einstein relation and excess entropy scaling law in liquid Copper]
{Stokes-Einstein relation and excess entropy scaling law in liquid Copper}
\author[N. Jakse, A. Pasturel]{N. Jakse, A. Pasturel}
\address{
Sciences et Ing\'{e}nierie des Mat\'{e}riaux et Proc\'{e}d\'{e}s, UMR CNRS 5266, Grenoble Universit\'{e} Alpes, \\BP 75, 38402 Saint-Martin d'H\`{e}res Cedex, France}

\date{Received September 8, 2015, in final form September 29, 2015}
\authorcopyright{N. Jakse, A. Pasturel, 2015}

\begin{document}

\maketitle

\begin{abstract}
We report an \emph{ab initio} study of structural and dynamic properties of liquid copper as a function of temperature. In particular, we have evaluated the temperature dependence of the self-diffusion coefficient from the velocity autocorrelation function as well the temperature dependence of the viscosity from the transverse current correlation function. We show that LDA based results are in close agreement with experimental data for both the self-diffusion coefficient and the viscosity over the temperature range investigated. Our findings are then used to test empirical approaches like the Stokes-Einstein relation and the excess entropy scaling law widely used in the literature. We show that the Stokes-Einstein relation is valid for the liquid phase and that the excess entropy scaling law proposed by Dzugutov is legitimate only if a self-consistent method for determining the packing fraction of the hard sphere reference liquid is used within the Carnahan-Starling approach to express the excess entropy.
\pacs 61.25.Mv, 61.20.Ja, 66.10.-x
\end{abstract}

\section{Introduction}

The study of the relationship between structural and dynamic properties of metallic liquids is a long-standing open problem in condensed matter physics. For instance, transport properties such as the shear viscosity ($\eta$) and the diffusion coefficients ($D$) of liquid metals are key parameters in the study of the crystal nucleation and growth in metallic melts \cite{CHE2011,TAN2012}.The temperature dependences of the viscosity and the relaxation time also play a very important role in studying the liquid-glass transition in a glass-forming system. Therefore, finding such a link should be also of practical interest as it relates a quantity like the self-diffusion coefficient that can be difficult to measure accurately to an experimentally more accessible dynamic, thermodynamic or structural quantity.

One alternative is to approximate diffusion coefficients from viscosities through the Stokes-Einstein (SE) relation \cite{ASS2010}, $D_\textrm{SE} = k_\textrm{B}T/2\pi R\eta$, where $k_\textrm{B}$ is Boltzmann's constant and $R$ is an effective diameter of particles. The interest of this relation lies on the fact that it is much more difficult to measure the diffusion coefficients of liquids than their viscosities since the influences of convective flow on the diffusion profile during annealing can be important using the long-capillary (LC) technique and its variation. At the same time, viscosity data are often available from standard data collections with a weak scatter between measurements made using different techniques. For instance, it is only $\pm 6\%$ for copper \cite{HAN2006}. First derived for the motion of a macroscopic particle in a homogeneous viscous medium, its use for atomic diffusion in liquid alloys requires defining $R$ at the atomistic scale. It is well accepted in the common literature \cite{ASS2010} to use the first position in the pair-correlation function obtained either experimentally or by means of atomistic simulations.

Recently, the accurate experimental determination of self-diffusion coefficients of species like Ni, Ti, or Cu \cite{MEY2008,HOR2009,MEY2010} from the use of quasielastic neutron scattering (QNS) has made it possible to extensively discuss the applicability of the Stokes-Einstein relation. The main result is that some experiments support its validity even for complex alloys \cite{MEY2002} while others evidence its failure \cite{BRI2001,BRI2008}, indicating the importance of the chemical composition of the melt. Moreover, it seems that the temperature also plays an important role in its use \cite{BRI2008}. These experiments raise many questions and among them the choice of the effective diameter of particles as well as in a more general aspect, the relation of transport properties with the melt structure.

Another alternative is to relate the dynamic properties to the structure and thermodynamics of melts using the universal scaling relationship like that proposed by Dzugutov \cite{DZU1996}. Within this scheme, the pair excess entropy of a liquid, $S_2$, is related to a dimensionless form of the diffusion coefficient, $D^*$. The quantity $S_2$ can be easily computed from pair correlation functions. However, if this approach is legitimate for model fluids like Lennard-Jones or Hard-Sphere (HS) fluids, a coupled Monte Carlo/molecular dynamics study has shown that this two-body correlation approximation does not hold for liquid metals and alloys modelled by EAM potentials \cite{HOY2000}.

More generally, molecular-dynamics (MD) simulations represent a powerful means for getting a microscopic picture of structural and dynamic properties in the liquid state and for studying their relationships but such studies are fruitful at the condition that atomic interactions are accurately modelled.

For liquid Cu, structural and dynamic properties have been extensively studied using the semi-empi\-rical embedded-atom model (EAM) formalism \cite{MEI1990,ALE1998,CHE2004,CEL2007,HAN2008,CHE2012}, as well as by the orbital free ab initio method \cite{BHU2012}. However, if structural properties are in very good agreement with experimental data whatever the EAM potential, diffusion coefficients reported in classical MD simulations show an important scatter, depending on the specific implementation of EAM potentials. Another difficulty holds with the fact that the calculated diffusion coefficients are mainly referred to experimental LC data \cite{HEN1961}, with the consequence that they exhibit a scatter of about $\pm 50\%$ around the most recent QNS values \cite{MEY2010}. Only simulated results provided by Alemany {et al.} \cite{ALE1998} disagree with LC data and these authors suggest that the LC values may be erroneous. Let us mention that very recent MD simulations \cite{CHE2012} refer to experimental QNS values and give a better agreement with them.

Within the density functional theory (DFT) framework, \emph{ab initio} MD (AIMD) simulations performed at 1500~K either by Pasquarello {et al.} \cite{PAS1992} or by Kresse and Hafner \cite{KRE1993} give calculated self-diffusion coefficients that differ by a factor of $2$. Whereas Pasquarello {et al.} reported a value of $(2.8\pm 0.2)\times10^{-9}$~m$^2$s$^{-1}$, Kresse and Hafner gave a value of $(5.6\pm 0.4)\times10^{-9}$~m$^2$s$^{-1}$. At the same temperature, the QNS value is $(4.33\pm 0.06)\times10^{-9}$~m$^2$s$^{-1}$ \cite{MEY2010}.

All these results show that it is necessary to revisit the determination of the structural and dynamic properties of liquid Cu using \emph{ab initio} molecular dynamics simulations.

In a first step, we show that the determination of the self-diffusion coefficient as well as the viscosity depend on the exchange-correlation functional used in DFT calculations, namely the local-density approximation (LDA) and the generalized gradient approximation (GGA). Our study shows that GGA based calculations of the diffusion coefficient underestimates both LDA based values and experimental data while we obtain the opposite trend for the viscosity. We evidence that these results can be explained by a more important cage effect using GGA as compared to LDA, due to the increase of the icosahedral motifs.

In a second step, our results are used to discuss the Stokes-Einstein relation and its validity as a function of temperature. Finally, we test the universal scaling law relating the diffusion coefficient and the excess entropy of a liquid as proposed by Dzugutov \cite{DZU1996}. We show that the scaling law is obeyed provided a self-consistent method for determining the packing fraction of the hard sphere reference fluid is used within the Carnahan-Starling approach to express the excess entropy. In order to strengthen the present discussion, we shall present our results together with those recently published for liquid Al \cite{JAK2013}.

\section{Simulation details}

The AIMD simulations have been performed by using the Vienna \emph{ab initio} simulation package \cite{KRE1996} using the projected augmented-wave method to describe the electron-ion interaction \cite{KRE1999}. A plane-wave basis set with an energy cutoff of 273~eV is used. The exchange-correlation energy is described using either the local-density approximation \cite{CEP1980a,CEP1980b} or the generalized gradient approximation in the Perdew-Burke-Ernzerhof form \cite{PER1996}. Only the $\Gamma$-point sampling is considered to represent the supercell Brillouin zone.

Equations of motion are solved using Verlet's algorithm in the velocity form with a time step of 1~fs. Atomic motions are carried out in the NVT ensemble by means of a Nose thermostat to control temperature. For each temperature, we use a cubic cell with periodic boundary conditions containing 256 atoms. The liquid sample is first equilibrated at 2000~K (well above the experimental melting temperature of 1358~K) and then cooled to 1800, 1600, and 1398~K, successively with a rate of $10^{13}$~K/s. At each temperature, we adjust the volume $V$ of the supercell to reproduce the experimental densities \cite{HAN2006} and after an equilibration of 10~ps, the run is continued further during 80~ps for production of the structural and dynamic quantities.

A number of $2000$ configurations are collected to calculate the static structure factor, the pair-correla\-tion function, as well as the diffusion coefficient and the viscosity. Among these configurations, we select ten independent ones, to compute their inherent structures \cite{STI1982} for the purpose of studying the local ordering. This is done numerically by carrying out a conjugate gradient energy-minimization in order to suppress the kinetic energy.

\section{Results and discussion}

\subsection{Self-diffusion coefficient}

\begin{figure}[!b]
\centerline{
\includegraphics[width=0.56\textwidth]{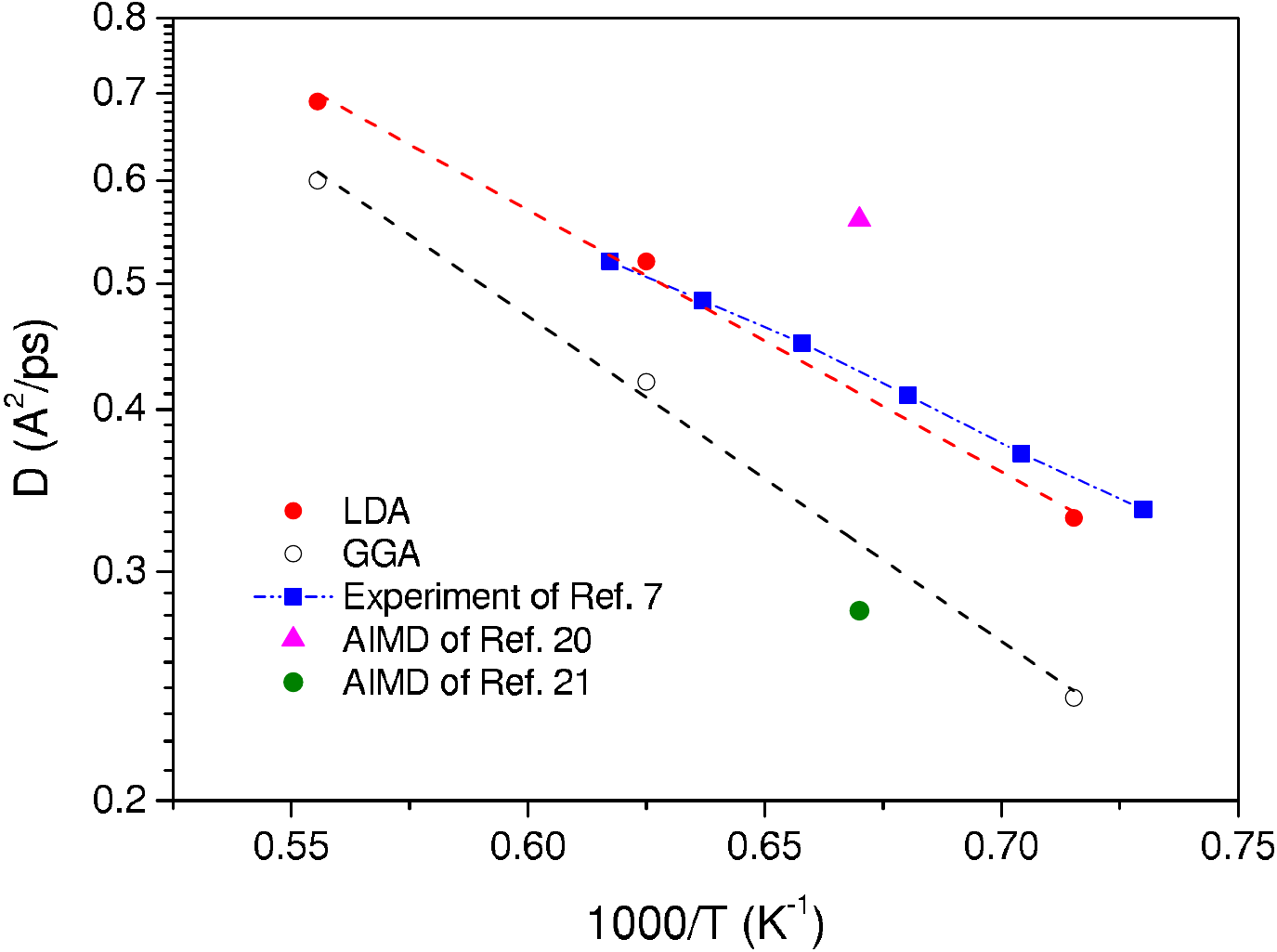}
}
\caption{(Color online) Evolution of the self-diffusion coefficient of liquid Cu as a function of inverse temperature. The solid circles are AIMD results with LDA and the open ones are those with GGA. The dashed lines are their respective Arrhenius fit. The full squares are experimental data of reference \cite{MEY2010} along with their Arrhenius fit (dash-dotted lines). AIMD results of references \cite{PAS1992} (full circle) and \cite{KRE1993} (triangle up) are also plotted for comparison.}
\label{Figure1}
\end{figure}

In a first step, we consider the single-atom dynamics through the velocity auto-correlation function, and its time integration that gives access to the self-diffusion coefficient, $D$ \cite{HAN2006}. Let us mention that this method requires shorter simulation times than that based on the mean-square displacement to have statistically meaningful results \cite{HAN2006,BRO1984}.

We report the temperature dependence of $D$ using the LDA and GGA approximations in figure~\ref{Figure1}. We can see that LDA calculations are close to the experimental QNS values \cite{MEY2010} while GGA calculations underestimate LDA calculations by about $20\%$ above the melting point. We also compute an activation energy assuming an Arrhenius-type behaviour for the diffusion process in liquid copper. As shown in figure~\ref{Figure1}, the data are well fitted in the whole range of temperatures and the derived activation energy is $E_D = 370 \pm 5$ and $475 \pm 5$~meV respectively for the LDA and GGA approximations. Let us mention that the LDA value is in close agreement with the experimental value, namely $337 \pm 5$~meV \cite{MEY2010}.

\subsection{Viscosity}

To compute the viscosity, we use a direct method based on the transverse current correlation function (see reference \cite{JAK2007} for a detailed description of this technique), which has the advantage of yielding a generalized $q$-dependent shear viscosity from which the hydrodynamic limit can be evaluated \cite{ALL1983,PAL1994}.

\begin{figure}[!h]
\centerline{
\includegraphics[width=0.56\textwidth]{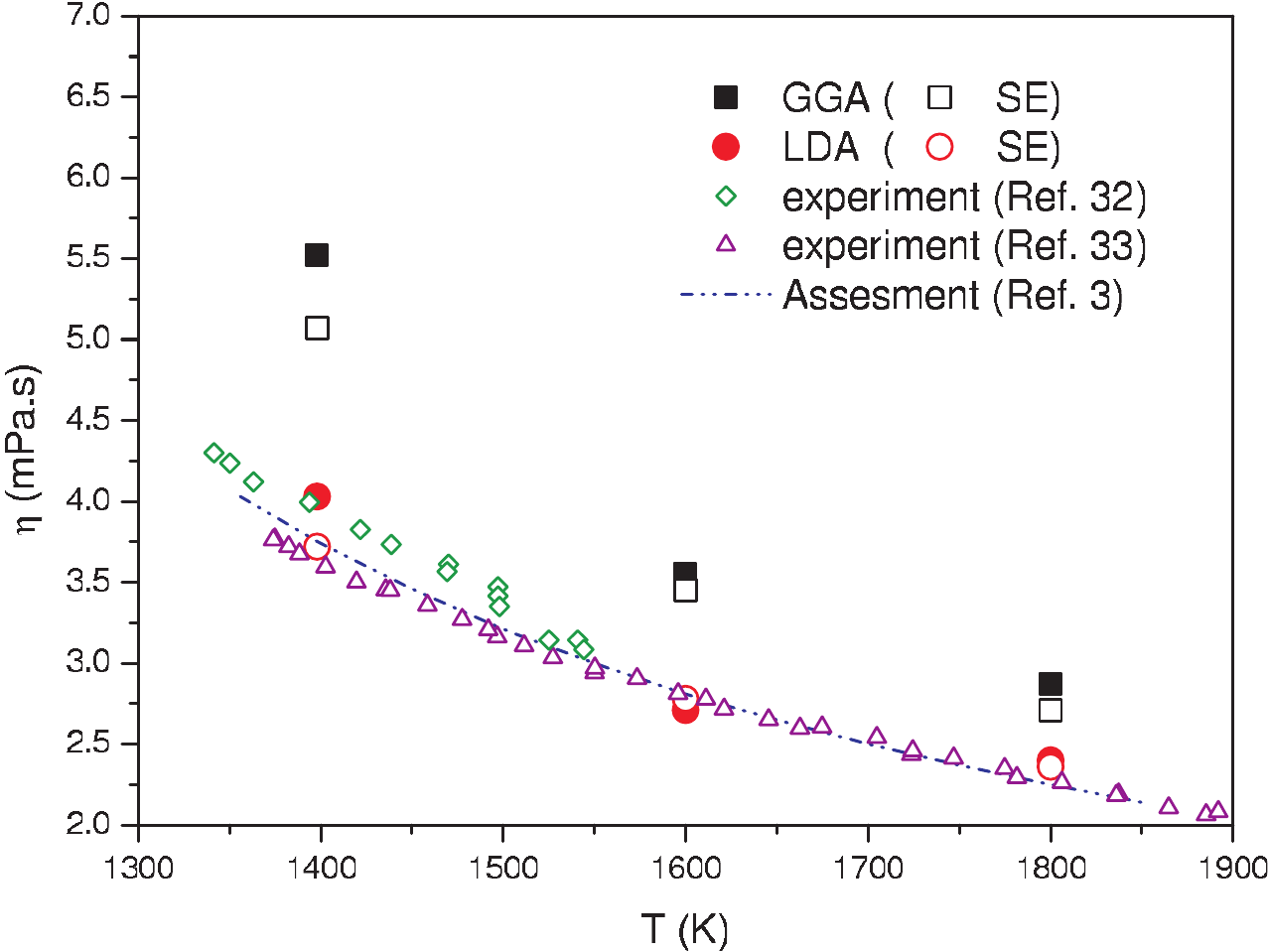}
}
\caption{(Color online) Viscosity of liquid Cu as a function of temperature. The full circles and full squares correspond to our AIMD results, respectively with LDA and GGA, as obtained with the transverse current-correlation functions. The open circles and open squares correspond to viscosities inferred from the self-diffusion coefficients, respectively with LDA and GGA, determined from our AIMD results using the Stokes-Einstein relation. The triangles, diamonds, and dash line correspond to the experimental values of references \cite{KEH2007} and \cite{BRI2007}, and the assessment of reference \cite{ASS2010}, respectively.}
\label{Figure2}
\end{figure}

In figure~\ref{Figure2}
we compare LDA and GGA results with the assessed values \cite{ASS2010}. As for the temperature dependence of the self-diffusion coefficient, we observe that the temperature dependence of the viscosity obtained by LDA is close to the temperature dependence of the assessed viscosity while GGA values overestimate LDA and the assessed values. More particularly, LDA values are consistent with experimental data of Kehr et al. \cite{KEH2007} and Brillo et al. \cite{BRI2007}, both experimental sets being used in the assessment. The derived activation energy assuming an Arrhenius-type behaviour is $E_\eta = 370 \pm 5$ and $475 \pm 5$~meV, respectively, for the LDA and GGA approximations.

Both computed self-diffusion coefficients and viscosities are then used to test the validity of the Stokes-Einstein relation. In figure~\ref{Figure2}, we report viscosities computed \emph{via} the Stokes-Einstein relation using $R$ given by the position of the first peak in the computed pair-correlation function. At $T = 1600$~K, the LDA value afforded by the Stokes-Einstein relation is 2.78~mPa$\cdot$s, close to the direct one, 2.71~mPa$\cdot$s and the assessed value of 2.81~mPa$\cdot$s. We obtain a similar agreement at higher temperature, i.e., $T = 1800$~K, while at lower temperature, namely $T = 1400$~K, we observe a small discrepancy, the value afforded by the Stokes-Einstein relation being $10\%$ smaller than the direct one. GGA calculations follow the same trend. For instance at $T = 1600$~K, the value obtained from the SE relation is 3.45~mPa$\cdot$s which can be compared to that obtained by the direct method, namely 3.55~mPa$\cdot$s. Note that the present study does not support the conclusions of the EAM based study \cite{CHE2012} claiming that the SE relation is not obeyed in liquid~Cu.

\subsection{Discussion}

In order to understand the significant discrepancies between GGA and LDA calculations of the self-diffusion coefficient and the viscosity, we inspect structural properties of liquid copper using both functionals.

\begin{figure}[!t]
\centerline{
\includegraphics[width=0.56\textwidth]{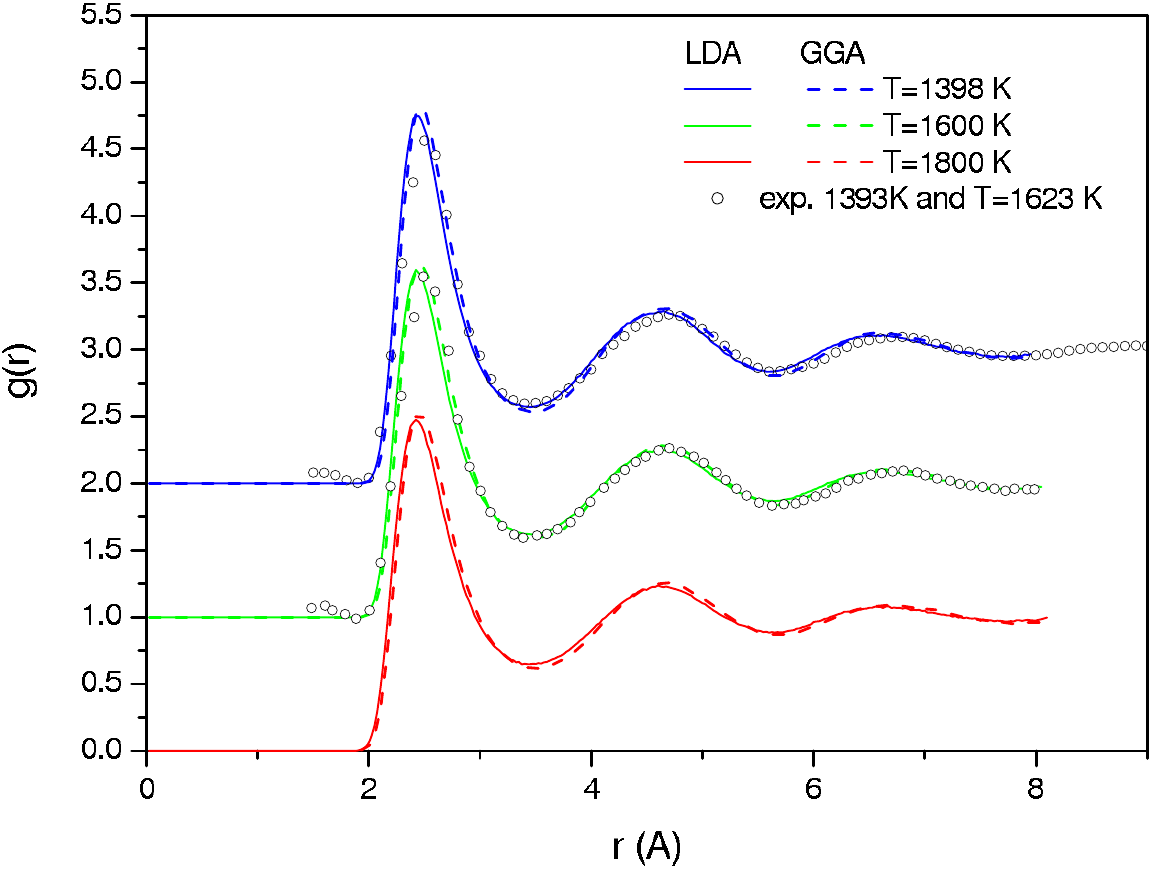}
}
\caption{(Color online) Evolution of the pair-correlation function with temperature. Open circles and triangles correspond to experimental values of references \cite{WAS1980} ($1393$~K) and \cite{EDE1980} ($1600$~K), respectively. The curves for $1600$~K, and $1398$~K are shifted upwards by an amount of 1 and 2, respectively.}
\label{Figure3}
\end{figure}

In figure~\ref{Figure3},
we report the pair-correlation function $g(r)$ for all the three temperatures in the liquid range using GGA and LDA. As already obtained for liquid aluminum \cite{JAK2013}, we do not see any appreciable difference between the two sets of calculations. Let us mention that our $g(r)$ at $T = 1600$~K compares well with $g(r)$ interpolated from X-ray diffraction experiments \cite{WAS1980}, while our $g(r)$ at $T = 1398$~K compare also well with the results from neutron diffraction experiments at $T = 1393$~K \cite{EDE1980}. The calculated $g(r)$ from previous \emph{ab initio} \cite{PAS1992,KRE1993} or EAM model \cite{MEI1990,ALE1998,CHE2004,CEL2007,HAN2008,CHE2012} are also consistent with our curves.

\begin{figure}[!b]
\centerline{
\includegraphics[width=0.56\textwidth]{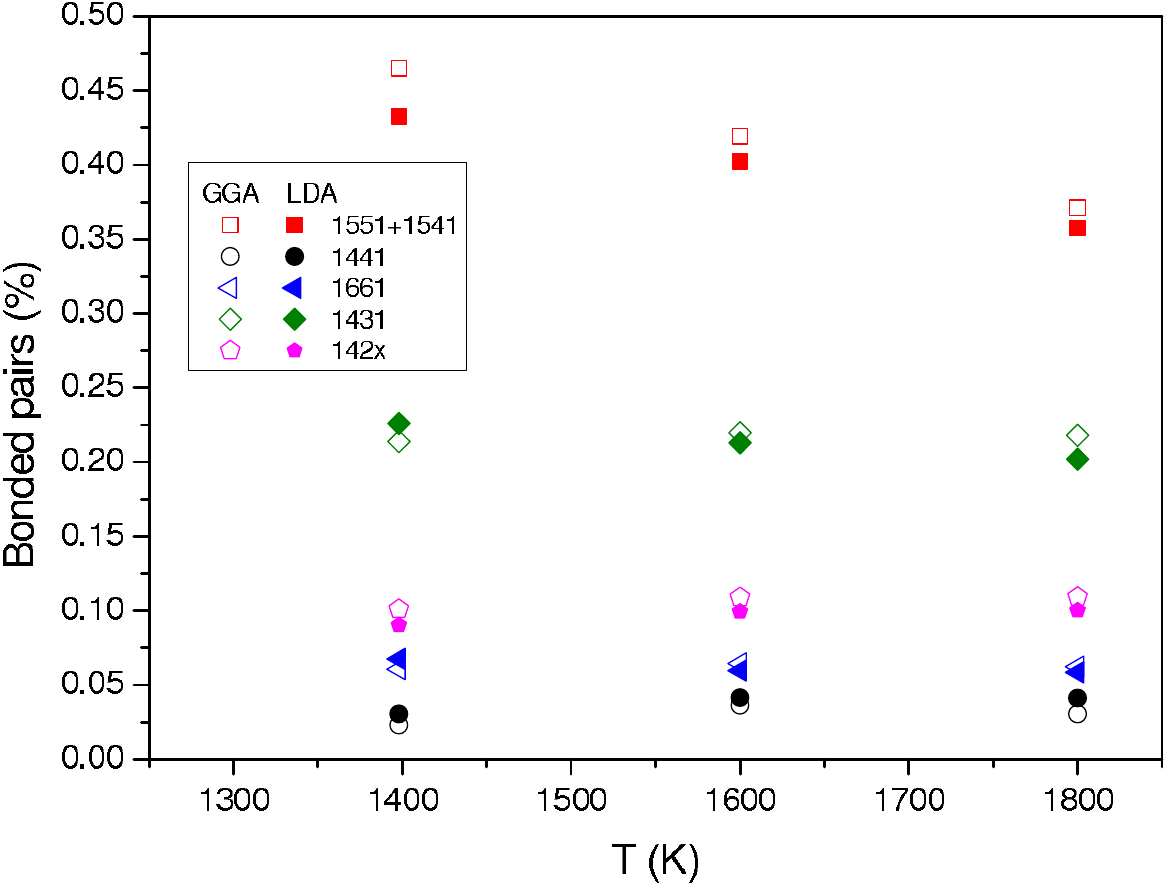}
}
\caption{(Color online) Evolution of abundances of the most important pairs for liquid Cu with temperature. The solid symbols are the results of LDA and the open ones are those of GGA. Error bars are typically of the order of $0.01$.}
\label{Figure4}
\end{figure}

Coordination numbers are obtained from the integration of the computed pair-correlation functions $g(r)$ up to its first minimum. We find an increase of coordination number from $N_\textrm{c} = 11.8\pm 0.1$ at $T = 1800$~K to $N_\textrm{c} = 12.4\pm 0.1$ at $T = 1398$~K while GGA calculations give $N_\textrm{c} = 12.0\pm 0.1$ at $T = 1800$~K and $N_\textrm{c} = 12.5\pm 0.1$ at $1398$ K. We obtain that LDA calculations give coordination numbers smaller than those obtained using GGA ones, but the differences remain small and cannot explain the differences obtained in the dynamic properties.

A deeper insight into the presence and the nature of the local atomic environments can be obtained from the common-neighbor analysis introduced in reference \cite{HON1987} and described in some detail in reference \cite{JAK2006}. In figure~\ref{Figure4},
we display the most abundant pairs, averaged over the ten inherent configurations for each temperature. For both LDA and GGA calculations, $142x$ (sum of $1422$ and $1421$), $1431$ and $15xx$ (sum of $1551$ and $1541$) pairs are found to be the main ones. We can also put forward two important remarks. 15xx pairs are the most abundant ones and they are the only pairs that display a linear increase as the temperature decreases. Let us mention that our results are in agreement with the previous \emph{ab initio} based analysis \cite{GAN2006,JAK2008}. The $15xx$ bonded pairs characterize the icosahedral order \cite{JAK2003-1,JAK2003-2,JAK2003-3} while the $142x$ bonded pairs are related to close packed structures like the face-centered cubic and hexagonal close-packed structures \cite{JAK2006}. The $1431$ pairs can be considered as distorted icosahedra or distorted close-packed structures \cite{PAS2010}. However, they cannot be responsible for the differences between dynamic properties using either GGA or LDA since they are similar in both approximations. The same comment holds for the $142x$ pairs. On the contrary, a significant increase of the number of $15xx$ pairs using the GGA approximation is a strong indication that the icosahedral short range order is more pronounced when using the GGA approximation.

\begin{figure}[!h]
\centerline{
\includegraphics[width=0.6\textwidth]{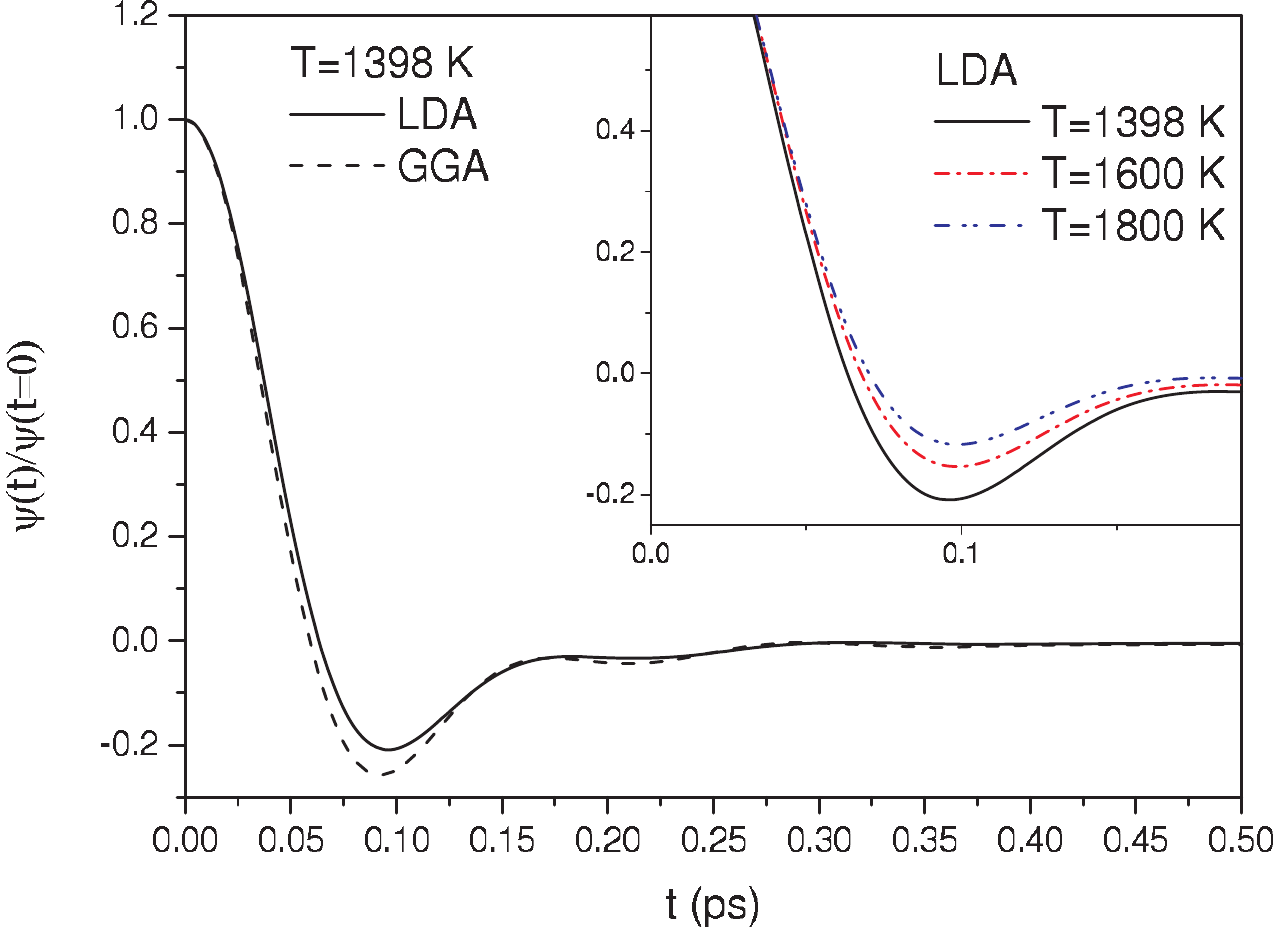}
}
\caption{(Color online) Velocity auto-correlation function obtained for liquid Cu at $1398$~K with LDA (solid lines) and GGA (dashed lines). Inset: Comparison in the vicinity of the main minimum with the LDA for $1398$~K, $1600$~K and $1800$~K.}
\label{Figure5}
\end{figure}

To study the influence of ISRO on the dynamic properties, we inspect the velocity auto-correlation function from which the self-diffusion coefficient is obtained. In figure~\ref{Figure5}
we report this function computed at $T = 1398$~K using the LDA and GGA approximations. The function decays quickly first and gets to a minimum value after nearly 0.1~ps, which is known as the cage effect due to the temporary trapping of atoms by their neighbours. We can note that this first minimum becomes deeper using GGA, which reveals an increase of the cage effect. This minimum is also less pronounced when the temperature increases, as shown in the inset for LDA (a similar behaviour is seen for the GGA). This behaviour can be related to the increase of the icosahedral motifs using GGA as compared to LDA for the same temperature or when the temperature decreases as seen in figure~\ref{Figure5}. Indeed, as discussed in reference \cite{JAK2008}, the backscattering regime, that is predominant for liquid Cu, is more pronounced in the phase which presents the highest ISRO and consequently the diffusivity is smaller.

\begin{figure}[!t]
\centerline{
\includegraphics[width=0.95\textwidth]{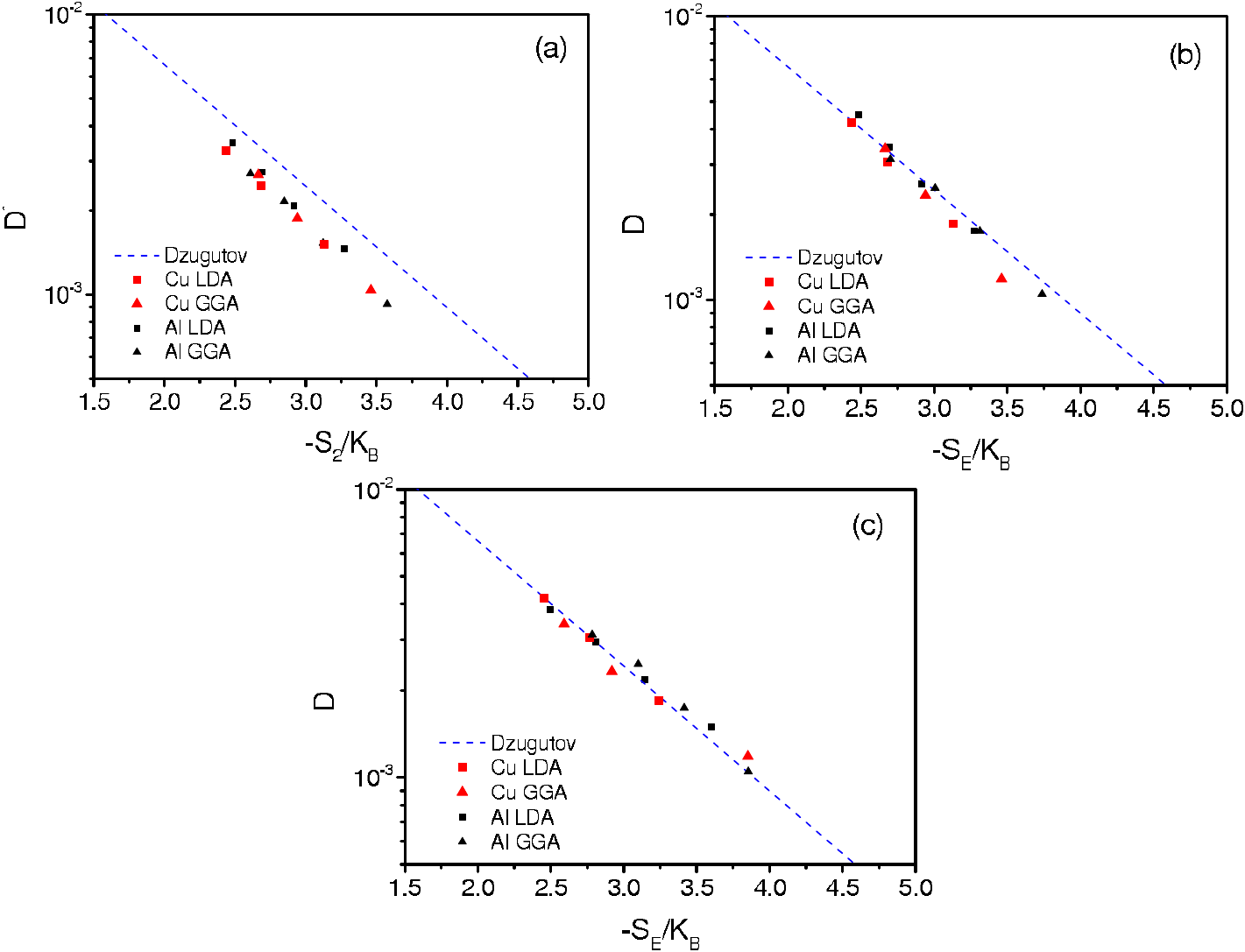}
}
\caption{(Color online) The scaled diffusion coefficient of Cu and Al (a) vs the $S_2$ approximation using \emph{ab initio} calculations, (b) vs the excess entropy calculated by the Carnahan Starling (CS) equation with the simple choice for $\sigma$ and $g(\sigma)$ (see text) as well as (c) with self-consistent values for $\sigma$ and $g(\sigma)$.}
\label{Figure6}
\end{figure}

We now discuss the universal scaling relationship proposed by Dzugutov \cite{DZU1996}, that relates a dimensionless form of the diffusion coefficient, $D^*$ to the excess entropy of the liquid phase, $S_\textrm{E}$. The relationship can be written as:
\begin{equation}
\label{EQ1}
D^*=0.049\exp(S_\textrm{E}),
\end{equation}
where $D^*$ is reduced with respect to the diffusion coefficient $D$ using uncorrelated binary collisions described by the Enskog theory as:
\begin{equation}
\label{EQ2}
D^*=D\Gamma^{-1}\sigma^{-2}.
\end{equation}
In equation~(\ref{EQ2}), $\Gamma$ is the Enskog collision frequency which can be calculated for the temperature $T$ and the number density $\rho$ within the framework of the hard-sphere fluid as:
\begin{equation}
\label{EQ3}
\Gamma=4\sigma^2 g(\sigma)\rho \left( \frac{\pi k_{\textrm{B}}T}{m}\right) ^{1/2},
\end{equation}
where $m$ and $\sigma$ are the atomic mass and the hard-sphere diameter. Parameter $g(\sigma)$ is the value of the HS pair-correlation function $g(r)$ at the contact distance. In a first
step, we study the scaling law assuming Dzugutov's idea that the excess entropy can be represented by the two-body correlation approximation denoted by:
\begin{equation}
\label{EQ4}
S_{2}=-2\pi \rho \int_{0}^{\infty }\left[ g(r)\ln g(r)-g(r)+1\right] r^{2}\rd r.
\end{equation}
Note that $\sigma$ and $g(\sigma)$ are approximated by the position and height of the first peak in the \emph{ab initio} computed pair-correlation function. Results for liquid Cu and previous ones for Al \cite{JAK2013} are displayed in figure~\ref{Figure6}~(a) and show that the $S_2$ approximation is also not sufficiently accurate when utilizing either LDA or GGA calculations. It is worth mentioning that the truncation errors in equation (\ref{EQ4}), due to the limited range of $g(r)$ calculated by AIMD, do not exceed $2\%$, and, therefore, do not affect the results for Dzugutov's law.

To go beyond the two-body correlation approximation, we use a more reliable determination of $S_\textrm{E}$ but still keeping the framework of the HS reference fluid. It is based on the Carnahan-Starling equation which is known to give a quasi-exact equation of states for HS fluids in a range of values of packing fraction characteristic of melts \cite{CAR1969}. In this case, the excess entropy can be written as:
\begin{equation}
\label{EQ5}
S_\textrm{E} = \frac{\xi(3\xi-4)}{(1-\xi)^2} -\frac{2(2-\xi)}{(1-\xi)^3}T\left(\frac{\partial\xi}{\partial T}\right)_V,
\end{equation}
where the packing density is given by $\xi = \pi\rho\sigma^3/6$. All entropy calculations can be done using the packing fraction $\xi$ and its evolution with temperature. To determine $\xi$, we use the experimental densities \cite{HAN2006} and $\sigma$ values obtained previously. Note that the temperature dependence of $\xi$ is mainly related to that of the density, as $\sigma$ does not vary in the investigated temperature range. From figure~\ref{Figure6}~(a), one can see that the agreement with the original fit of Dzugutov is better since the scatter in data becomes smaller.

To explain the remaining discrepancy, we turn back to the determination of $\sigma$ and $g(\sigma)$. It is well known from perturbation theories \cite{HAN2006} that the effective HS diameter adjusted to represent the structure and thermodynamics of a real system is different from the simple choice for $\sigma$ and $g(\sigma)$ made above. In particular, this method leads to a lack of thermodynamic consistency between the HS reference fluid and the real one. To evidence this failure, we compare in table~\ref{Table1} the isothermal compressibility calculated by the Carnahan Starling expression, i.e., $\rho k_\textrm{B} T\chi_T=(1-\xi)^4/[2\xi(4-\xi)+(1-\xi)^4]$ (values in parenthesis) with that obtained from the extrapolation of $S(q)$ at $q = 0$, $S(q)$ being the structure factor obtained from \emph{ab initio} calculations. Table~\ref{Table1} indicates for both Al and Cu that the thermodynamic consistency, i.e. $\rho k_\textrm{B} T\chi_T=S(0)$, \emph{via} calculations of the isothermal compressibility is not respected. We present only LDA calculations for comparison as we obtain the same trend for GGA based results.

\begin{table}[htb]
\caption{Values of the packing fraction $\xi$ and isothermal compressibility $\chi_T$ as a function of temperature $T$ for Cu and Al (reference \cite{JAK2013}) obtained from the self-consistent procedure using the integral equation method (LDA calculations). The values in parenthesis correspond to the simple choice for $\sigma$ and $g(\sigma)$. The last column corresponds to \emph{ab initio values} (see text). The typical error bars on the AIMD results are~0.005.} \label{Table1} \vspace{2ex}
\begin{center}
\renewcommand{\arraystretch}{0}
\begin{tabular}{|c|c||c|c|c|}
\hline\hline
 & $T$ (K) & $\xi =\pi \rho \sigma ^{3}/6$ & $\frac{(1-\xi )^{4}}{2\xi (4-\xi)+(1-\xi )^{4}}$ & $\rho k_{\textrm{B}}\, T\chi _{T}=S(0)$ \\
\hline
&&&&\\
\hline
\raisebox{-1.7ex}[0pt][0pt]{Cu}
 & 1398 & 0.461 (0.565) & 0.027 (0.0090) & 0.030\strut \\
\cline{2-5}
 & 1600 & 0.434 (0.545) & 0.034 (0.0112) & 0.035\strut \\
\cline{2-5}
 & 1800 & 0.416 (0.534) & 0.039 (0.0125) & 0.041\strut \\
\hline
\raisebox{-1.7ex}[0pt][0pt]{Al}
 & 875 & 0.466 (0.563) & 0.024 (0.0093) & 0.029\strut \\
\cline{2-5}
& 1000 & 0.446 (0.553) & 0.029 (0.0103) & 0.031\strut \\
\cline{2-5}
 & 1125 & 0.431 (0.547) & 0.033 (0.0109) & 0.034\strut \\
\cline{2-5}
 & 1250 & 0.415 (0.538) & 0.038 (0.0119) & 0.040\strut \\
\hline\hline
\end{tabular} \renewcommand{\arraystretch}{1}
\end{center}
\end{table}

Therefore, to enforce the thermodynamic consistency between the HS fluid and the liquid described from AIMD, we impose that the pair-correlation entropy of the hard sphere model $S_2^\textrm{HS}$, is equal to the pair-correlation entropy provided by LDA or GGA calculations, respectively $S_2^\textrm{LDA}$ and $S_2^\textrm{GGA}$. To obtain $S_2^\textrm{HS}$, we compute the pair-correlation function of the hard sphere model, $g^\textrm{HS} (r)$, using accurate integral equation method \cite{JAK2003b}. This method is based on the exact Ornstein-Zernike convolution equation
\begin{equation}
\label{EQ6}
h(r)=c(r)+\rho \int h(r^{\prime })c(| \mathbf{r}-\mathbf{r}^{\prime }| )\rd\mathbf{r}^{\prime }=c(r)+\gamma(r),
\end{equation}
where $h(r)$, $c(r)$ and $\gamma(r)$ are respectively the total, direct and indirect correlation functions between two atoms, separated by a distance $r$ in a liquid composed of $N$ atoms with number density $\rho$ \cite{HAN2006}. In order to determine the pair-correlation
function $g(r) = h(r) + 1$, equation~(\ref{EQ6}) should be solved together with a closure relation whose formal expression is as follows:
\begin{equation}
\label{EQ7}
h(r)=\exp \left[ -\beta u^\textrm{HS}(r)+\gamma (r)+B(r)\right] -1,
\end{equation}
In equation~(\ref{EQ7}), $B(r)$ is the so-called bridge function composed of an infinite series of elementary diagrams \cite{HAN2006}, and $u^\textrm{HS} (r)$ is the hard-sphere potential. For $B(r)$ we use the efficient approximate formulation for the hard-sphere model proposed by Rodgers and Young \cite{ROG1984}:
\begin{equation}
\label{EQ8}
B(r)=-\gamma (r) + \ln \left[ 1+ \frac{A\gamma (r)-1}{A}\right],
\end{equation}
which ensures an accurate description of thermodynamic properties of the HS fluid through the optimization of parameter $A$ \cite{BRE1992}. The set of equations~(\ref{EQ6}), (\ref{EQ7}), and (\ref{EQ8}) are solved numerically using the algorithm of Labik et al. \cite{LAB1985}, which combines the Newton-Raphson method and the traditional iterative technique. Following the authors of reference~\cite{CHA2001,CHA2001a}, we use the tangent linear differentiation technique that is an essential ingredient for improving the accuracy of the integral equation method and for computing thermodynamic functions of the HS fluid that involve the derivative of $g(r)$.

From table~\ref{Table1}, we can note that our approach leads to similar values of the isothermal compressibility calculated using either the Carnahan Starling expression, i.e. $\rho k_\textrm{B} T\chi_T=(1-\xi)^4/[2\xi(4-\xi)+(1-\xi)^4]$, or the extrapolation of  $S(q)$ at $q = 0$, i.e. $\rho k_\textrm{B} T\chi_T=S(0)$, $S(q)$ being obtained from LDA. This close correspondence is a strong indication of the thermodynamic consistency of our approach to determine the excess entropy.

Taking advantage of this simple analytical expression of the excess entropy, we plot in figure~\ref{Figure6}~(c) the relationship between the excess entropy and the dimensionless form of the diffusion coefficient, $D^*$ for both Cu and Al data. It is found that the scaling law proposed by Dzugutov is legitimate with the excess entropy derived from the Carnahan-Starling approach. Let us mention that this approach does not require any adjustable parameter.

\section{Conclusion}

In summary, we have computed the temperature dependence of the self-diffusion coefficient as well the viscosity of liquid copper by \emph{ab initio} molecular dynamics using two different exchange and correlation potentials, LDA and GGA. The comparison of the calculated self-diffusion coefficients with the most recent QNS experimental data and the comparison of the calculated viscosities with the assessed values favor the LDA approximation to compute the dynamic properties of liquid copper. We show that the origin of the discrepancy using GGA is due to an enhancement of the icosahedral short range order (ISRO) that directly impact the dynamic properties of liquid copper. Our results allow us to discuss the applicability of the SE relation to the description of the relation between diffusivity and viscosity and to the firm establishment of its validity for liquid copper over the temperature range investigated. Finally, we show that the proposed scaling law of Dzugutov which relates the scaled diffusivity to the excess entropy can be  successfully applied if a self-consistent method for determining the pair-correlation function of the HS fluid is used as well as the Carnahan Straling approach for expressing the excess energy is introduced. Note that the excess entropy based on the Carnahan Starling equation is given by an analytical expression and does not contain any adjustable parameter. How this formulation can be applied to liquids such as silicon, which presents loosely packed structures, is an important open question to be addressed in a future work.

\section*{Acknowledgements}
We acknowledge the CINES and IDRIS under Project Number INP2227/72914 as well as PHYNUM CIMENT for computational resources. This work was performed within the framework of the Centre of Excellence of Multifunctional Architectured Materials ``CEMAM'' number ANR-10-LABX-44-01 funded by the ``Investments for the Future'' Program.

\clearpage

\ukrainianpart

\title{Співвідношення Стокса-Ейнштейна і  скейлінговий закон для надлишкової ентропії у рідкій міді}
\author{Н. Жакс, А. Пастурель}
\address{
Наука і техніка матеріалів і процесів,
UMR CNRS 5266, Альпійський університет,  Гренобль, Франція}

\makeukrtitle

\begin{abstract}
У цій статті представлено  \emph{ab initio} дослідження структурних і динамічних властивостей рідкої міді як функції температури. Зокрема,
проведено розрахунок температурної залежності коефіцієнта самодифузії з автокореляційної функції швидкостей, а також температурної залежності швидкості з поперечної кореляційної функції струму. Показано, що результати, базовані на  наближенні локальної густини, добре узгоджуються з експериментальними даними як для коефіцієнта дифузії, так і для в'язкості в межах досліджуваного діапазона температур. Ці результати використані
для перевірки емпіричних підходів типу співвідношення Стокса-Ейнштейна та скейлінгового закону для надлишкової ентропії, які широко використовуються в літературі.  Показано, що співвідношення Стокса-Ейнштейна задовільняється для рідкої фази і скейлінговий закон для надлишкової ентропії,  запропонований Дзугутовим, є легітимним лише за умови використання самоузгодженого методу для визначення фракції упакування  твердосферної рідини, коли  підхід Карнагана-Старлінга використано для отримання виразу  надлишкової ентропії.
\keywords рідка мідь, співвідношення Стокса-Ейнштейна, універсальні скейлінгові закони, {ab initio} молекулярна динаміка
\end{abstract}


\begin{thebibliography}{99}

\bibitem{CHE2011} Cheng Y.Q., Ma E.,
Prog. Mater. Sci., 2011, \textbf{56}, 379; \bibdoi{0.1016/j.pmatsci.2010.12.002}.

\bibitem{TAN2012} Tanaka H.,
Eur. Phys. J. E, 2012, \textbf{35}, 113; \bibdoi{10.1140/epje/i2012-12113-y}.

\bibitem{ASS2010}  Assael M.J., Kalyva A.E., Antoniadis K.D., Banish R.M., Egry~I., Wu~J., Kaschnitz~E., Wakeham~W.A.,
J. Phys. Chem. Ref. Data, 2010, \textbf{39}, 033105; \bibdoi{10.1063/1.3467496}.

\bibitem{HAN2006} Hansen J.P., McDonald I.R., Theory of Simple Liquids, 3rd Edn.,
Elsevier Inc., Amsterdam, 2006.

\bibitem{MEY2008} Meyer A., Stober S., Holland-Moritz D., Heinen O., Unruch T.,
Phys. Rev. B, 2008, \textbf{77}, 092201; \\\bibdoi{10.1103/PhysRevB.77.092201}.

\bibitem{HOR2009} Horbach J., Rozas R.E., Unruch T., Meyer A.,
Phys. Rev. B, 2009, \textbf{80}, 212203; \bibdoi{10.1103/PhysRevB.80.212203}.

\bibitem{MEY2010}  Meyer A.,
 Phys. Rev. B, 2010, \textbf{81}, 012102; \bibdoi{10.1103/PhysRevB.81.012102}.

\bibitem{MEY2002} Meyer A.,
Phys. Rev. B, 2002, \textbf{66}, 134205; \bibdoi{10.1103/PhysRevB.66.134205}.

\bibitem{BRI2001} Brillo J., Pommrich I., Meyer A.,
Phys. Rev. Lett., 2001, \textbf{107}, 165902; \bibdoi{10.1103/PhysRevLett.107.165902}.

\bibitem{BRI2008} Brillo J., Chathoth S.M., Koza M.M., Meyer A.,
Appl. Phys. Lett., 2008, \textbf{93}, 121905; \bibdoi{10.1063/1.2977863}.

\bibitem{DZU1996} Dzugutov M.,
Nature, 1996, \textbf{381}, 137; \bibdoi{ 10.1038/381137a0}.

\bibitem{HOY2000} Hoyt J.J., Asta M., Sadigh B.,
Phys. Rev. Lett., 2000, \textbf{85}, 594; \bibdoi{10.1103/PhysRevLett.85.594}.

\bibitem{MEI1990} Mei J., Davenport J.W.,
Phys. Rev. B, 1990, \textbf{42}, 9682; \bibdoi{10.1103/PhysRevB.42.9682}.

\bibitem{ALE1998} Alemany M.M.G., Rey C., Gallego L.J.,
J. Chem. Phys., 1998, \textbf{109}, 5175; \bibdoi{10.1063/1.477133}.

\bibitem{CHE2004} Chen F.F., Zhnag H.F., Qin F.X., Hu Z.Q.,
J. Chem. Phys., 2004, \textbf{120}, 1826; \bibdoi{10.1063/1.1636452}.

\bibitem{CEL2007} Celino M., Rosato V., Di Cicco A., Trapananti A., Massobrio C.,
Phys. Rev. B, 2007, \textbf{75}, 174210; \\ \bibdoi{10.1103/PhysRevB.75.174210}.

\bibitem{HAN2008} Han X.J., Chen M., L\"u Y.J.,
Int. J. Thermophys., 2008, \textbf{29}, 1408; \bibdoi{10.1007/s10765-008-0489-7}.

\bibitem{CHE2012} Lu Y., Cheng H., Chen M.,
J. Phys. Chem., 2012, \textbf{136}, 214505; \bibdoi{10.1063/1.4723683}.

\bibitem{BHU2012} Bhuiyan G.M., Gonz\'{a}lez L.E., Gonz\'{a}lez D.J.,
Condens. Matter Phys., 2012, \textbf{15}, 33604; \bibdoi{10.5488/CMP.15.33604}.

\bibitem{HEN1961} Henderson J., Yang L.,
Trans. Metall. Soc. AIME, 1961, \textbf{221}, 72.

\bibitem{PAS1992} Pasquarello A., Laasonen K., Car R., Lee C., Vanderbilt~D.,
Phys. Rev. Lett., 1992, \textbf{69}, 1982; \\ \bibdoi{10.1103/PhysRevLett.69.1982}.

\bibitem{KRE1993} Kresse G., Hafner J.,
Phys. Rev. B, 1993,  \textbf{48}, 13115; \bibdoi{10.1103/PhysRevB.48.13115}.

\bibitem{JAK2013} Jakse N., Pasturel A.,
Sci. Rep., 2013, \textbf{3}, 3135; \bibdoi{10.1038/srep03135}.

\bibitem{KRE1996} Kresse G., Furthm\"uller J.,
Comput. Mater. Sci., 1996,  \textbf{6}, 15; \bibdoi{10.1016/0927-0256(96)00008-0}.

\bibitem{KRE1999} Kresse G., Joubert D.,
Phys. Rev. B, 1999, \textbf{59}, 1758; \bibdoi{10.1103/PhysRevB.59.1758}.

\bibitem{CEP1980a} Ceperley D.M., Alder B.J.,
Phys. Rev. Lett., 1980, \textbf{45}, 566; \bibdoi{10.1103/PhysRevLett.45.566}.

\bibitem{CEP1980b}
Perdew J.P., Zunger A.,
Phys. Rev. B, 1981, \textbf{23}, 5048; \bibdoi{10.1103/PhysRevB.23.5048}.

\bibitem{PER1996} Perdew J.P., Burke K., Ernzerhof M.,
Phys. Rev. Lett., 1996, \textbf{77}, 3865; \bibdoi{10.1103/PhysRevLett.77.3865}.

\bibitem{STI1982} Stillinger F.H., Weber T.A.,
Phys. Rev. A, 1982, \textbf{25}, 978; \bibdoi{10.1103/PhysRevA.25.978}.

\bibitem{BRO1984} Brown D., Clarke J.H.R.,
Mol. Phys., 1984, \textbf{51}, 1243; \bibdoi{10.1080/00268978400100801}.

\bibitem{JAK2007} Jakse N., Wax J.F., Pasturel A.,
J. Chem. Phys., 2007, \textbf{126}, 234508; \bibdoi{10.1063/1.2741521}.

\bibitem{ALL1983} Alley W.E., Alder B.J.,
Phys. Rev. A, 1983, \textbf{27}, 3158; \bibdoi{10.1103/PhysRevA.27.3158}.

\bibitem{PAL1994} Palmer B.J.,
Phys. Rev. E, 1994, \textbf{49}, 359; \bibdoi{10.1103/PhysRevE.49.359}.

\bibitem{KEH2007}  Kehr M., Hoyer W., Egry I.,
Int. J. Thermophys., 2007, \textbf{28}, 1017; \bibdoi{10.1007/s10765-007-0216-9}.

\bibitem{BRI2007}  Brillo J., Brooks R., Egry I., Quested P.N.,
Int. J. Mater. Res., 2007, \textbf{98}, 457; \bibdoi{10.3139/146.101494}.

\bibitem{WAS1980} Waseda Y., The Structure of Non-Crystalline Materials, McGraw-Hill, New York, 1980.

\bibitem{EDE1980} Eder O.J., Erdpresser E., Kunsch B., Stiller H., Suda M.J.,
J.~Phys.~F: Met. Phys., 1980, \textbf{10}, 183; \\ \bibdoi{10.1088/0305-4608/10/2/008}.

\bibitem{HON1987} Honeycutt J.D., Andersen H.C., J. Phys. Chem., 1987, \textbf{91}, 4950; \bibdoi{10.1021/j100303a014}.

\bibitem{JAK2006} Jakse N., Pasturel A.,
Mod. Phys. Lett. B, 2006, \textbf{20}, 655; \bibdoi{10.1142/S0217984906011177}.

\bibitem{GAN2006} Ganesh P., Widom M.,
Phys. Rev. B, 2006, \textbf{74}, 134205; \bibdoi{10.1103/PhysRevB.74.134205}.

\bibitem{JAK2008} Jakse N., Pasturel A.,
Phys. Rev. B, 2008, \textbf{78}, 214204; \bibdoi{10.1103/PhysRevB.78.214204}.

\bibitem{JAK2003-1} Jakse N., Pasturel A.,
Phys. Rev. Lett., 2003, \textbf{91}, 195501; \bibdoi{10.1103/PhysRevLett.91.195501}.

\bibitem{JAK2003-2}
Jakse N., Le Bacq O., Pasturel A.,
Phys. Rev. B, 2004, \textbf{70}, 174203; \bibdoi{10.1103/PhysRevB.70.174203}.

\bibitem{JAK2003-3} Jakse N., Pasturel A.,
J. Chem. Phys., 2004, \textbf{120}, 6124; \bibdoi{10.1103/PhysRevLett.93.207801}.

\bibitem{PAS2010} Pasturel A., Tasci E.S., Sluiter M.H.F., Jakse N.,
Phys. Rev. B, 2010, \textbf{81}, 140202R; \bibdoi{10.1103/PhysRevB.81.140202}.

\bibitem{CAR1969} Carnahan N.F., Starling K.E.,
J. Chem. Phys., 1969, \textbf{51}, 635 (1969); \bibdoi{10.1063/1.1672048}.

\bibitem{JAK2003b} Jakse N., Charpentier I.,
Phys. Rev. E, 2003, \textbf{67}, 061203; \bibdoi{10.1103/PhysRevE.67.061203}.

\bibitem{ROG1984} Rogers F.J., Young D.A.,
Phys. Rev. A, 1984, \textbf{30}, 999; \bibdoi{10.1103/PhysRevA.30.999}.

\bibitem{BRE1992} Bretonnet J.L., Jakse N.,
Phys. Rev. B, 1992, \textbf{46}, 5717; \bibdoi{10.1103/PhysRevB.46.5717}.

\bibitem{LAB1985} Labik S., Malijevski A., Vonka P.,
Mol. Phys., 1985, \textbf{56}, 709;  \bibdoi{10.1080/00268978500102651}.

\bibitem{CHA2001} Charpentier I., Jakse N.,
J. Chem. Phys., 2001, \textbf{114}, 2284; \bibdoi{ 10.1063/1.1332808}.

\bibitem{CHA2001a}
Charpentier I., Jakse N.,
J. Chem. Phys., 2005, \textbf{123}, 204910; \bibdoi{10.1063/1.2117010}.

\end{thebibliography}
\end{document}